\documentclass[pra,%
 reprint,
 amsmath,assume,
 aps,
]{revtex4-2}
\usepackage{mathrsfs} 
\usepackage{graphicx}
\usepackage{dcolumn}
\usepackage{lipsum}
\usepackage{bm}
\usepackage[usenames]{color}
\usepackage{colortbl}
\usepackage{hyperref}
\usepackage{natbib}
\begin{document}

\title{{\em In-situ} tunable superconducting diode: towards field-free operation with infinite nonreciprocity}

\author{Razmik A. Hovhannisyan}
\author{Taras Golod}
\author{Amirreza Lotfian}
\author{Vladimir M. Krasnov}
\email{vladimir.krasnov@fysik.su.se}
\affiliation{ Department of Physics, Stockholm University, AlbaNova University Center, SE-10691 Stockholm, Sweden} 

\date{\today}
\begin{abstract}
Efficient, scalable, and magnetic-field-free superconducting diodes are essential for future superconducting electronics; yet, despite significant efforts, such practical devices remain unrealized. The main challenge lies in achieving broad-range {\em in-situ} tunability, both for optimization and for achieving transistor-like operation.
Here, we study diodes based on four-terminal niobium planar Josephson junctions. We show that the multiterminal structure eliminates the need for an external magnetic field and enables essentially unrestricted \textit{in-situ} tunability, along with reconfigurability of the diode polarity, leading to new functionality. For example, we demonstrate that such diodes can operate as Gauss neurons via reentrant superconductivity. 
By deliberately tuning the junction parameters, we obtain effectively infinite nonreciprocity (within experimental resolution) leading to threshold-free ac-current rectification. Such technologically simple, reconfigurable, and broadly tunable diodes could be instrumental for future digital and neuromorphic computing.
\end{abstract}

\maketitle

\newpage

\section*{Introduction}

The semiconducting field-effect transistor (FET) is historically the most successful electronic component driving the ongoing digital revolution. The key advantage of the FET lies in its extreme {\em in-situ} tunability, with an on/off ratio of $\sim 10^4$--$10^{6}$ or higher, enabling effective signal isolation at VLSI and ULSI integration levels. However, power dissipation in semiconducting circuits has led to a surge in energy consumption by rapidly growing data centers, posing serious sustainability challenges and motivating the development of innovative post-CMOS solutions with improved energy efficiency \cite{IEA}.

The growing demand for digital data capacity has renewed interest in the development of classical superconducting computers \cite{Likharev_1991,Holmes_2013,Ortlepp_2014,Golod_2015,Soloviev_2017,Tolpygo_2019,Mukhanov_2019}, which offer the potential for drastic improvements in both speed and power efficiency. However, a major challenge remains the lack of efficient, scalable, and technologically simple superconducting analogues of key semiconductor components such as diodes and transistors.

The diode is one of the simplest building blocks in semiconductor electronics, with diverse applications in signal processing and digital circuits. In recent years, the development of superconducting diodes (ScDs) has become an active area of research \cite{Krasnov_1997,Ono_2022,Banerjee_2024,Nadeem_2023,Nagaosa_2024,Ando_2020,Wu_2022,Lin_2022,Baumgartner_2022,Golod_2022,Gupta_2023,Finkelstein_2023,Sundaresh_2023,Guarcello_2024,Deskhmukh_2024,Giazotto_2025,Giazotto_2023,Berggren_2025,Vilegas_2003,Moschalkov_2006,Lustikova_2018,Koelle_2000,Robinson_2023,Beck_2005,Goldobin_2024}. The key figures of merit for ScDs are the nonreciprocity of the critical current, $A = \mid I_c^+ / I_c^- \mid$. 

The emergence of nonreciprocity requires simultaneous breaking of both spatial and time-reversal symmetries. Spatial symmetry can be broken by utilizing noncentrosymmetric superconductors~\cite{Nagaosa_2024, Ando_2020, Wu_2022, Lin_2022, Nadeem_2023, Banerjee_2024} and heterostructures~\cite{Ono_2022, Baumgartner_2022, Sundaresh_2023}, or using geometrical asymmetry in vortex ratchets~\cite{Vilegas_2003, Moschalkov_2006, Lustikova_2018, Koelle_2000, Robinson_2023}, nanowires \cite{Giazotto_2023,Giazotto_2025,Berggren_2025}, and Josephson junctions (JJs)~\cite{Krasnov_1997, Golod_2022, Beck_2005,Guarcello_2024,Goldobin_2024}.
The geometrical approach allows the use of conventional superconductors \cite{Golod_2022}, making it suitable for practical applications.  

The time reversal symmetry can be broken by application of magnetic field. However, external magnetic field is unacceptable in complex superconducting circuits. Consequently, efforts are underway to develop field-free ScDs \cite{Wu_2022,Golod_2022,Ono_2022,Finkelstein_2023}. 

An ScD with infinite nonreciprocity could address one of the main challenges in superconducting electronics — the lack of efficient signal isolation. 
However, maximum nonreciprocity is achieved only under specific conditions, and {\em in-situ} adjustment is required to reach this regime. Simple and efficient tunability would enable both optimization of diode performance and transistor-like control of the current--voltage ($I$--$V$) characteristics, opening a wide range of potential applications. Thus, similarly to the case of FETs, broad-range {\em in-situ} tunability is essential for the practical success of ScDs.

In this work, we study ScDs based on four-terminal planar Nb JJs. We aim to achieve maximal nonreciprocity without an external magnetic field. To this end, we first clarify the conditions for optimal operation and subsequently tune the ScDs to this point. Such deliberate engineering has enabled the realization of a near-perfect diode with $A > 1000$. 
By exploiting the multiterminal structure of our JJs, we introduce novel split-current operation modes, which provide essentially unlimited \textit{in-situ} tunability, eliminate the need for an external magnetic field, and facilitate new functionalities such as Gauss neuron operation with reentrant superconductivity.

\section*{Results}

We study four-terminal variable-thickness-bridge type JJs made of a thin Nb film.
Figures~\ref{fig:1}(a–c) show a sketch and scanning electron microscope (SEM) images of two (D1 and D2) of more than ten studied ScDs. 
D1 contains a JJ with a relatively large $J_c$ and has wide electrodes ($W_z = 3.86~\mu\text{m}$) and a shorter junction ($W_x = 4~\mu\text{m}$). 
In contrast, a JJ in D2 has a lower $J_c$, much narrower electrodes, $W_z = 0.48~\mu\text{m}$, and larger length, $W_x = 5~\mu\text{m}$. 
The four electrodes form an X-shape patten. The two electrodes at each side of the JJ are separated by a narrow ($\sim 100~$nm) notch. 

Characteristics of all devices are listed in the Supplementary. The fabrication procedure and measurement setup are described in Methods and Supplementary.

\subsection*{Self-field inductance}

Nonreciprocity in our JJs is induced by the self-field effect -- the back-action of the bias current on the critical current \cite{Krasnov_1997,Monaco_2012,Vasenko_1981,Guarcello_2024}. The phenomenon is caused by a current-induced Josephson phase gradient, $\nabla \varphi$, which is related to the phase gradient in the superconducting electrode, $\nabla \varphi_s$, via the 
second London equation,
\begin{equation}
    \frac{\Phi_0}{2\pi}{\bf\nabla} \varphi_s = -{\bf A} - \mu_0 \lambda_L^2 {\bf J_s}.
\end{equation}
Here, $\Phi_0$ is the flux quantum, ${\bf A}$ is the vector potential, $\lambda_L$ is the London penetration depth, and ${\bf J_s}$ is the supercurrent density.

At least three factors contribute to the self-field effect.

\begin{figure*}[!ht]
    \begin{center}
    \includegraphics[width = 0.95\textwidth]{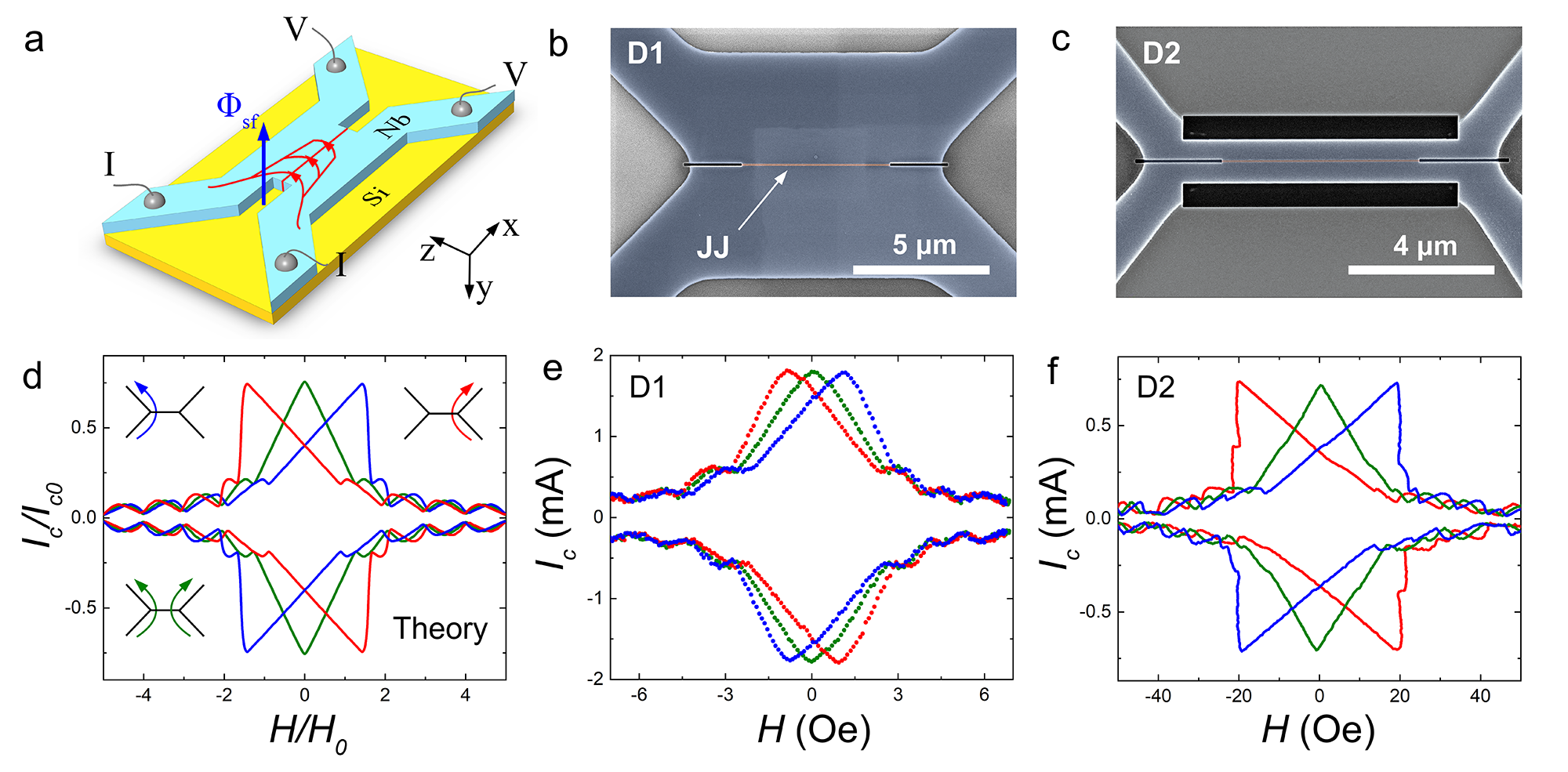}
\caption{\textbf{Geometry-dependent and bias-reconfigurable diodes.}
\textbf{a} A sketch of four-terminal Josephson diode. Asymmetric application of a bias current at one edge of the junction induces an effective self-flux $\Phi_{sf}$ back-acting on the critical current. 
\textbf{b} and \textbf{c} SEM images (false colour) of D1 and D2 devices with wide and narrow electrodes. Planar junctions are marked by orange lines. 
\textbf{d} Calculated $I_{c}(H)$ for a symmetric (olive) and asymmetric bias from left (blue) or right (red) edges. The self-field effect tilts the $I_c(H)$ patterns and induces nonreciprocity at finite field. \textbf{e} and \textbf{f} Measured $I_c(H)$ of D1 and D2 at $T=5$ K for three bias configurations as in \textbf{d}. Note that the tilt of $I_c(H)$ is much larger for D2, despite a smaller $I_c$. 
}
    \label{fig:1}
    \end{center}
\end{figure*}

The first contribution, the ${\bf A}$-term in Eq.~(1), gives rise to a field-induced Josephson phase gradient \cite{Krasnov_1997,Vasenko_1981,Monaco_2012,Krasnov_2020,Guarcello_2024},
\begin{equation}
    \frac{\partial \varphi}{\partial x} = \frac{2\pi}{W_x} \frac{H_y}{H_0},
\end{equation}
where $H_0$ is the flux-quantization field and $W_x$ is the JJ width. In the considered planar geometry, $H_y$ represents the out-of-plane field component. The current-induced self-field depends on the electrode geometry and can be enhanced by a narrow notch at the JJ edge, as illustrated in Fig.~\ref{fig:1}(a). This ``true'' self-field effect can be described by the geometric inductance of the electrodes, $L_g$.

The second contribution, associated with the ${\bf J_s}$ term in Eq.~(1), is related to the kinetic inductance, $L_k$. 
Since $\partial \varphi_s/\partial x$ is determined by the $x$-component of ${\bf J_s}$ along the JJ, $L_k$ is enhanced by counterflowing currents in the two electrodes along the junction, similarly to the geometric inductance in the notch. However, in contrast to $L_g$, $L_k$ is not associated with a real magnetic field.

The third contribution originates from a non-uniform distribution of the bias current density, $J_b(x)$, and the Josephson critical current density, $J_{c}(x)$, or other JJ parameters \cite{Vasenko_1981,Krasnov_1997}. Such non-uniformity induces a phase gradient via the dc Josephson relation $J_{c}(x)\sin[\varphi(x)] = J_b(x)$, which may appear unrelated to magnetic field. More precisely, it does originate from the self-field, but in the perpendicular in-plane, in our case, direction \cite{Krasnov_2020}.

Thus, the effective self-field inductance has both geometric and kinetic contributions,
\begin{equation}
    L_{sf}=L_g+L_k.
\end{equation}

\subsection*{Reconfigurable nonreciprocity}

As seen from Figs.~\ref{fig:1}(a–c), our diodes are spatially symmetric. The symmetry breaking is introduced through asymmetric biasing \cite{Krasnov_1997,Golod_2019,Golod_2022}. 
{The effective self-flux, $\Phi_{sf} = L_{sf} I_b^*$, is determined by the asymmetric part of the bias current, $I_b^*$. } 

Figure~\ref{fig:1}(d) shows the calculated 
$I_c(H)$ modulation for the symmetric (olive) and two asymmetric bias configurations from left (blue) and right (red) edges. 
It can be seen that a symmetric bias 
leads to a conventional reciprocal $I_c(H)$. In contrast, asymmetric biasing leads to the appearance of self-field, which tilts the $I_c(H)$ patterns. The relative displacement of the $I_c^+(H)$ and $I_c^-(H)$ branches leads to nonreciprocity, $I_c^+ \ne |I_c^-|$, and induces the ScD effect at finite field \cite{Krasnov_1997}. 

The remarkable feature of our diodes is their reconfigurability~\cite{Golod_2022}. The diode polarity can be flipped either by changing the bias configuration or the sign of the magnetic field. 
Figs.~\ref{fig:1}(e) and (f) show the measured $I_c(H)$ patterns for D1 and D2 at $T = 5$~K for different bias configurations. Apparently, the self-field effect in D2 is much more pronounced than in D1, demonstrating that the magnitude of the self-field effect—and thus the nonreciprocity—strongly depends on the electrode geometry and junction parameters. 

\begin{figure*}[t]
    \begin{center}
    \includegraphics[width = 0.99\textwidth]{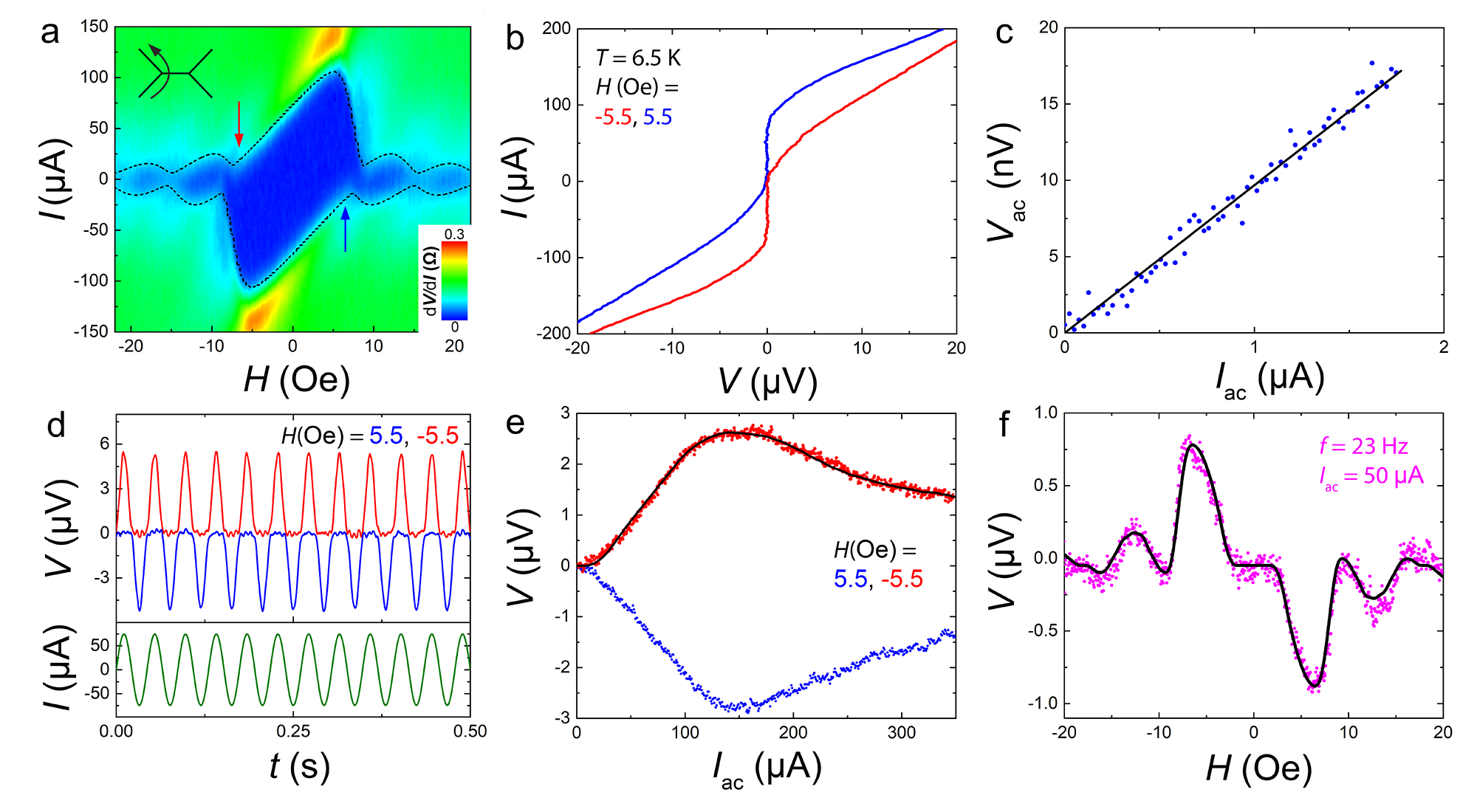}
    {\caption{\textbf{Threshold-free rectification of thermally tuned near-perfect diode D2. }
\textbf{a} The measured $I_c(H)$ modulation (colour map of differential resistance) in the left-edge bias configuration at $T=6.5$ K. The black dotted line represents a numerical fit for $W_x=3\lambda_J$. \textbf{b} The $I$-$V$s at $H = \pm 5.5$~Oe, indicated by arrows in (a). 
Note that the critical current is significant, $\sim 100~\mu$A, in one direction, but vanishing in the opposite direction. 
\textbf{c} High-resolution lock-in measurements at $H=-5.5$ Oe. The threshold-free Ohmic behavior confirms the absence of the critical current. 
\textbf{d} Time dependence of the ac transport current at $f = 23$~Hz (bottom) and instantaneous voltages measured at magnetic fields of 5.5 Oe (blue) and -5.5 Oe (red).
\textbf{e} Rectified dc-voltage as a function of ac-current amplitude for $H=\pm 5.5$ Oe. 
\textbf{f} Magnetic field dependence of the rectified dc-voltage (symbols) at $I_{\mathrm{ac}} = 50~\mu$A. Black lines in (e) and (f) represent numerical calculations.
}
\label{fig:2} }

    \end{center}
\end{figure*}

\subsection*{Optimal conditions}

The largest nonreciprocity is achieved when the maximum of $I_c(H)$ on one side is aligned with the minimum on the other side. The central maximum, $I_{c0}$, occurs at $\Phi=0$ and, therefore, appears at a field needed to compensate for the self-flux. The first minimum occurs at 
$\Phi = \Phi_0$. Since $I_c(\Phi_0) \simeq 0$, at the minimum the self-field is negligible and the flux is created predominantly by the external field. Thus, the condition for achieving maximum nonreciprocity is
\begin{equation}
    \Phi_{sf} = L_{sf} I_{c0} = \Phi_0.
\end{equation}

Both parameters in Eq.~(4) are tunable. $L_{sf}$ depends on the electrode geometry near or within the JJ (see Supplementary Material), while $I_{c0} = J_{c} W_x$ increases with the JJ length and the critical current density. However, a constraint exists. Achieving $A = \infty$ requires $I_c(\Phi_0) = 0$, which is possible only in the short-junction limit~\cite{Vasenko_1981},
\begin{equation}
W_x < 4\lambda_J,
\end{equation}
where $\lambda_J(J_{c})$ is the Josephson penetration depth. Therefore, Eq.~(5) imposes upper limits on both $W_x$ and $J_{c}$.

Optimization of ScD can be done both by proper geometric design ($L_{sf}$, $W_{x,z}$) and by adjustment of the junction parameters ($I_{c0}$) to satisfy Eqs.~(4) and (5). However, the former alternative requires physical modification of the device, while the latter enables broad {\em in-situ} tunability via temperature dependence, $J_{c}(T)$.

\subsection*{Tuning by temperature}

As can be seen from Fig.~\ref{fig:1}(f), D2 at $T=5$~K is characterized by a large self-field with the central lobe displacement by more than $3H_0$. However, the nonreciprocity is modest, $A \simeq 7$, for two reasons: the maxima are not aligned with the opposing minima, and the minima do not vanish because the JJ is in the long limit, as indicated by the broad triangular central lobe of $I_c(H)$~\cite{Vasenko_1981}.   

Fine-tuning of D2 was performed by raising the temperature. Figure~\ref{fig:2}(a) shows the $I_c(H)$ pattern at $T \simeq 6.5$~K. The $I_{c0}$ was reduced, bringing the junction into the short limit, $W_x = 3\lambda_J$, as follows from the numerical fit (black line). Importantly, the central maxima at this $T$ is aligned with the opposing first minima. 

Fig.~\ref{fig:2}(b) 
shows the $I$–$V$s at the fields of maximum nonreciprocity, $H = \pm 5.5$~Oe. It is seen that $I_c$ is significant $\mid I_c\text{(max)}\mid \simeq 100~\mu$A) in one direction but undetectable in the opposite direction. 
To evaluate the precision of the $I_c\text{(min)}$ cancellation, we performed accurate lock-in measurements, presented in Fig.~\ref{fig:2}(c). 
The offset-free Ohmic behaviour indicates that $\mid I_c(\text{min})\mid =0$ within our experimental resolution ($\sim 0.1~\mu$A).  
Thus, conscious optimization led to the realization of a perfect (within the resolution) ScD with $A > 10^3$. 

\subsection*{Threshold-free rectification}

Signal processing by ac-signal rectification is one of the main diode applications. 
Figs.~\ref{fig:2}(d-f) summarize rectification of an ac-transport current by D2. Fig. \ref{fig:2}(d) shows the sinusoidal current wave 
and instantaneous voltages measured at $H=\pm 5.5$ Oe. 
The current amplitude, $I_{ac}=75~\mu\text{A} < I_c\text{(max)}\simeq 100~\mu\text{A}$. Therefore, the voltage appears only at one half of the period, either negative for $H=5.5$~Oe, or positive for $H=-5.5$~Oe. Importantly, the optimized diode exhibits threshold-free rectification, as shown in Fig. \ref{fig:2}(c), in contrast to unoptimized diodes with finite nonreciprocity 
\cite{Golod_2022}.

Figure~\ref{fig:2}(e) shows the dependence of rectified dc-voltage on the ac-current amplitude. The dc voltage first increases with increasing $I_{ac}$, however at $I_{ac}>I_c\text{(max)}$ it starts to decrease due to appearance of the opposite signal at the second half of the period \cite{Golod_2022}. 

Fig.~\ref{fig:2}(f) shows magnetic field dependence of the rectified voltage at $I_{ac}=50~\mu$A. The oscillatory $V_{dc}(H)$ dependence reflects the shape of $I_c(H)$ modulation, Fig. \ref{fig:2}a, and the related oscillation of nonreciprocity \cite{Krasnov_1997,Golod_2022}. 
The black lines in Figs.~\ref{fig:2}(e) and (f) represent numerical modelling using the actual JJ parameters (see Methods), in good agreement with the experimental data.

\begin{figure*}[!ht]
    \begin{center}
    \includegraphics[width = 0.95\textwidth]{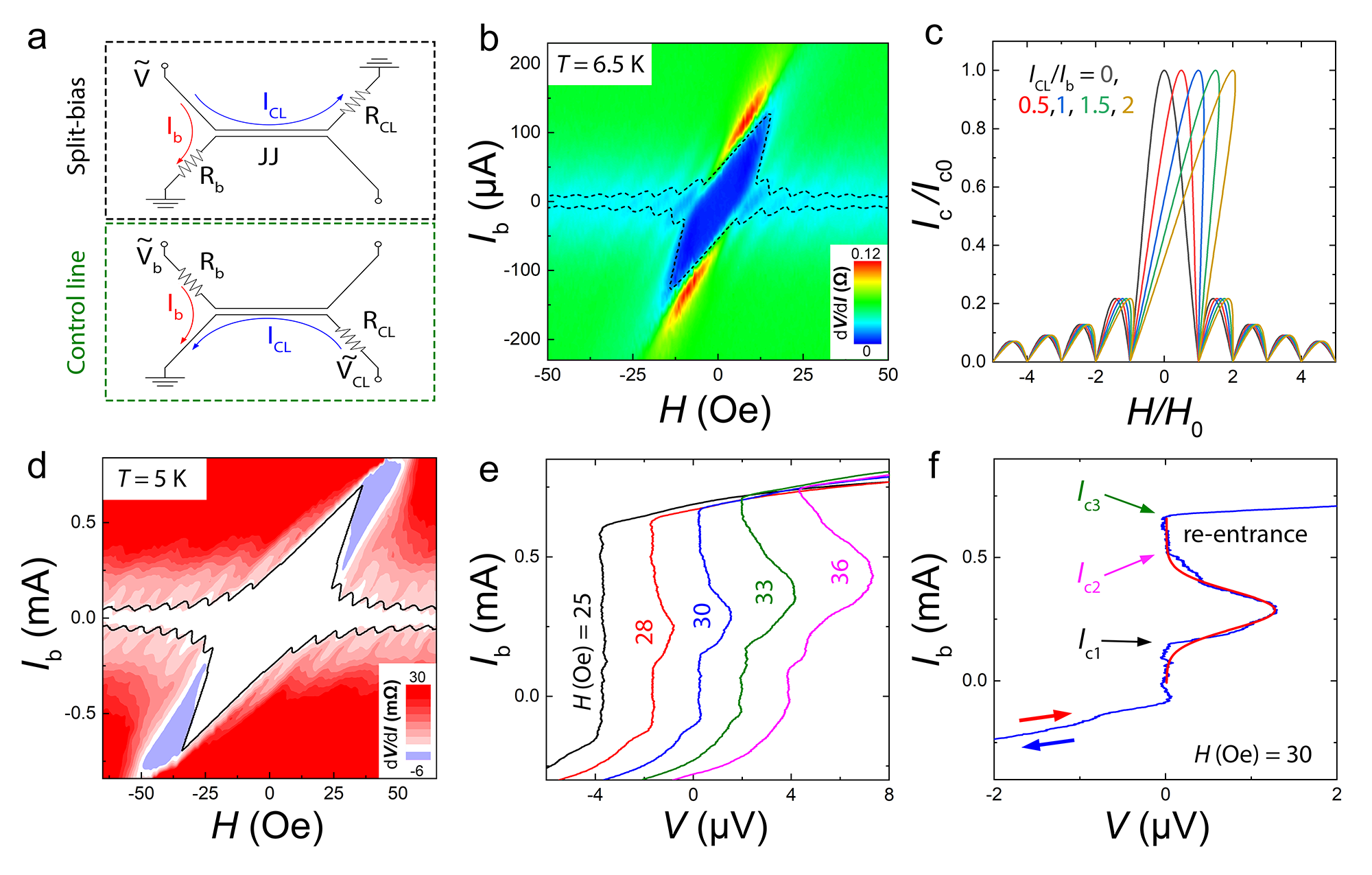}
\caption{
\textbf{Split-current operation: unrestricted tunability and reentrant superconductivity.}
\textbf{a} Schematics of the split-current and control-line operation modes. 
\textbf{b} The measured $I_c(H)$ colour map of D2 at $T=6.5$~K for $I_{\mathrm{CL}}/I_b = 1$. The dashed black line shows the numerically modelled $I_c(H)$ pattern.
\textbf{c} Calculated $I_c(H)$ patterns for a short junction in the split-current mode. Note appearance of over-tilting at two largest split ratios.  
\textbf{d} The measured $I_c(H)$ colour map of D2 at $T=5$~K for $I_{\mathrm{CL}}/I_b = 1$. The self-flux becomes so large that the central lobe of $I_c(H)$ overlaps with five higher-order lobes. The blue area marks the region of negative differential resistance. The black line represents the calculated $I_c(H)$ for $W_x/\lambda_J = 8$. 
\textbf{e} Evolution of the current--voltage characteristics in the region of negative differential resistance from (d). 
\textbf{f} Demonstration of reentrant superconductivity upon increasing bias current and Gauss-neuron operation. The blue line shows the $I$-$V$ at $H=30$~Oe. 
The red line represents the Gaussian fit.
\label{fig:3}
}
    \end{center}
\end{figure*}

\subsection*{Split-current operation mode}

The nonreciprocity of our diodes is determined by the tilt of $I_c(H)$ patterns. As can be seen from Figs. \ref{fig:1} (d) and (f), proper engineering enables practically vertical $I_c(H)$ slopes. 
However, according to earlier studies \cite{Vasenko_1981,Monaco_2012,Guarcello_2024}, the self-field effect can not over-tilt $I_c(H)$ more than that, i.e. can not change the sign of the slope $dI_c/dH$. This restriction is removed in our X-shaped JJs.

The four-terminal JJ structure enables sending currents in two arms simultaneously \cite{Monaco_2012,Guarcello_2024}, as sketched in the top panel of Figure~\ref{fig:3}(a). A bias part, $I_b$, passes through the JJ, while the other part, $I_{\mathrm{CL}}$, flows along the electrode acting as the control line (CL). 

Figure~\ref{fig:3}(b) shows $I_c(H)$ 
for D2 at $T=6.5$~K in the split-current mode with $I_{\mathrm{CL}}/I_b=1$. It can be seen that the tilt of the $I_c(H)$ pattern doubled compared to the ordinary bias mode at the same conditions, Fig. 2(a). 

The extent of CL-current contribution to the self-field effect is determined by the split ratio $I_{\mathrm{CL}}/I_b$. Fig.~\ref{fig:3}(c) shows numerically simulated $I_c(H)$ patterns for several split ratios. Simulations are made for a short JJ with a symmetric bias $I_b(x)=\text{const}$, which leads to a reciprocal Fraunhofer modulation in the absence of $I_{\text{CL}}$ (black line). The addition of $I_{\mathrm{CL}}$ tilts the $I_c(H)$ patterns in proportion to $I_{\mathrm{CL}}/I_b$. Since this ratio is unrestricted, the diode tunability is almost unlimited. In particular, it becomes possible to surpass the restriction on the sign change of $dI_c/dH$ achieving an overlap of the central lobe with subsequent higher-order lobes. 

Fig.~\ref{fig:3}(d) shows a measured $I_c(H)$ 
for D2 at lower $T = 5\,\mathrm{K}$ with $I_{\mathrm{CL}}/I_b=1$. In contrast to the corresponding pattern at $I_{\mathrm{CL}}/I_b=0$ shown in Fig.~\ref{fig:1}(f), here the central lobe is over-tilted so much that it overlaps with five subsequent lobes. Fig.~\ref{fig:3}(e) demonstrates the development of the $I$--$V$'s in the overlap region. It is seen that a profound nonmonotonous behaviour emerges. 

In Fig.~\ref{fig:3}(f) we focus on the $I$-$V$ at $H=30$ Oe. With increasing current, the JJ first switches to the resistive state at $I_{c1} \simeq 150~\mu\mathrm{A}$ corresponding to the fourth lobe in the $I_c(H)$ pattern from Fig. ~\ref{fig:3} (d). With further increase of the current, the voltage  decreases back to zero at $I_{c2} \simeq 500~\mu\mathrm{A}$, which corresponds to the lower edge of the over-tilted central lobe. Emergence of such reentrant superconductivity is accompanied by the appearance of negative differential resistance, marked by blue colour in Fig. 3 (d). Finally, above $I_{c3}\simeq 700~\mu$A, corresponding to the upper edge of the central lobe, the sample goes in the ordinary (monotonous) resistive state. 

\begin{figure*}[t]
    \begin{center}
    \includegraphics[width = 0.95\textwidth]{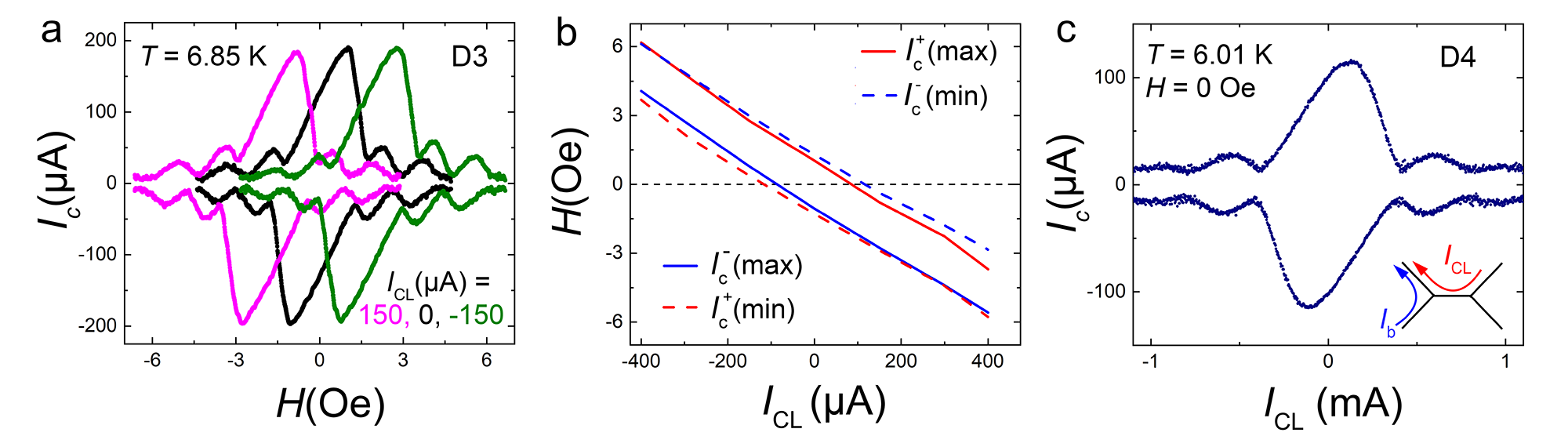}
    { \caption{
\textbf{Control-line operation mode without an external magnetic field.}
\textbf{a} Measured $I_c(H)$ modulation for D3, at different control line currents applied along one of junction electrodes, $I_{\mathrm{CL}} = 150, 0, -150~\mu\mathrm{A}$ (left, middle, right curves).
\textbf{b} Field positions of the main maxima (solid lines) and the first minima (dashed lines) of $I_c(H)$ as a function of $I_{\mathrm{CL}}$ for D3. The horizontal dashed line indicates that the points of maximum nonreciprocity occur at zero external field for $I_{\mathrm{CL}}\simeq \pm 100~\mu$A. 
\textbf{c} Measured critical current as a function of the control-line current $I_{\mathrm{CL}}$ for a device D4 at zero external field. The inset shows the sketch of current flow. The maximum nonreciprocity without external magnetic field is achieved at $I_{\mathrm{CL}}\simeq \pm 340~\mu$A.}    
\label{fig:4}
}
    \end{center}
\end{figure*}

\subsection*{Control line operation mode}

An external magnetic field is undesirable in complex superconducting circuits. However, it is often required for device operation. In superconducting digital electronics, the need for an external field is avoided by employing control lines 
\cite{Likharev_1991,Tolpygo_2019, Mukhanov_2019}.
The multiterminal geometry of our device enables CL operation, using JJ electrodes, as sketched in the bottom panel of Fig. 3(a). 

Figure~\ref{fig:4}(a) shows measured $I_c(H)$ patterns without (middle curve) and with constant $I_{\mathrm{CL}}= \pm 150~\mu\text{A}$ (left and right curves) for a diode D3. It is seen that the CL current simply shifts the $I_c(H)$ patterns without changing their shape. Fig.~\ref{fig:4}(b) shows the fields of the central positive and negative maxima, $I_c^+$(max) and $I_c^-$(max), as well as the first positive and negative minima, $I_c^+$(min) and $I_c^-$(min), as a function of $I_{\mathrm{CL}}$. 
It can be seen that the CL current creates an effective local field in the JJ $\sim 15$~Oe per 1 mA. The horizontal dashed line in Fig.~\ref{fig:4}(b) indicates that the points of maximum nonreciprocity are brought to $H = 0$ at $I_{\mathrm{CL}} \simeq \pm 100~\mu\text{A}$ for this device. 

Figure~\ref{fig:4}(c) shows the measured $I_c$ vs. $I_{\text{CL}}$ at zero external field for D4. 
The maximum nonreciprocity at $H = 0$ for this device is achieved at $I_{CL} = \pm 150~\mu\text{A}$. Thus, an external magnetic field is not required for the operation of our diodes. Due to the multiterminal configuration, they can be comfortably operated at $H = 0$ using small control line currents on the order of $100~\mu\text{A}$.

\section*{Discussion}

The studied here ScDs are technologically and conceptually simple. 
They are based on conventional low-$T_c$ Nb JJs and rely on geometric symmetry breaking, which can be precisely controlled and tailored to achieve the desired characteristics. 
The specified conditions for optimal operation, Eqs.~(4) and (5), together with quantitative estimates of self-field inductances presented in the Supplementary Material, enable the rational design and engineering of such diodes. 

A characteristic feature of our diodes is their inherent reconfigurability enabled by the multiterminal structure. As shown in Fig.~\ref{fig:1}(f), the diode polarity can be readily reversed by changing the bias configuration. Such reconfigurability could be instrumental for building advanced electronic components \cite{Golod_2022,Giazotto_2023}, including programmable logic gates, switchable signal routers, and multiplexers. 

Two strategies for achieving infinite nonreciprocity have been presented. The first is based on careful geometric engineering, with final fine-tuning via temperature adjustment (Fig.~2). This approach is suitable for signal routing without leakage to other channels. 

The second approach, representing the central novelty of this work, lies in the introduction of split-current and control-line operation modes. The split-current mode allows essentially unlimited {in-situ} tunability (Fig.~\ref{fig:3}), while the CL mode enables operation without an external magnetic field (Fig.~\ref{fig:4}). 
Both modes exploit the four-terminal structure of our JJs, which allows a portion of the current to flow along the JJ electrode, acting as a control line, as sketched in Fig.~\ref{fig:3}(a). 

The difference between the modes lies in the value and control of $I_{\mathrm{CL}}$.
In the split-current mode, a single current source is connected to one of the four terminals, while two other terminals are grounded via series resistors $R_{\mathrm{CL}}$ and $R_b$. In this configuration, the split-current ratio remains constant, $I_{\mathrm{CL}}/I_b = R_b/R_{\mathrm{CL}}$. In contrast, the CL mode employs two independent current sources, such that $I_{\mathrm{CL}}$ is fixed, while the ratio $I_{\mathrm{CL}}/I_b$ is variable.

The CL contribution to the self-flux is,
\begin{equation}
    \Phi_{sf} = L_{\text{JJ}} I_b + L_{\text{CL}} I_{\text{CL}} = L_{\text{JJ}} I_b \left[ 1 + \frac{L_{\text{CL}}}{L_{\text{JJ}}} \frac{I_{\text{CL}}}{I_b} \right],\label{L_CL}
\end{equation}
where $L_{\text{JJ}}$ and $L_{\text{CL}}$ are the self-field inductances of the JJ and the CL, respectively. In the conventional bias mode considered so far \cite{Golod_2022,Goldobin_2024}, only the first term is present. As follows from Eq.~(\ref{L_CL}), the split-current mode enables broad tunability via the unrestricted $I_{\text{CL}}/I_b$. 

In principle, any diode can be brought to the optimal operating point, Eq.~(4), without physical modification by selecting an appropriate split ratio. However, this broad tunability comes at the expense of reduced sensitivity. Since $I_{\text{CL}}$ does not pass through the JJ, it does not contribute to the voltage response and may, for example, reduce the rectification efficiency.

Therefore, the split-current mode should not be regarded as a universal approach for transforming an imperfect device into an ideal diode. Rather, it serves as a tool for fine-tuning a reasonably well-performing diode. For example, it can be used to eliminate a small residual threshold current. 
In this case, the gain in sensitivity due to the removal of the threshold can outweigh the loss of sensitivity caused by current splitting.

The purpose of the CL operation mode is to eliminate the need for an external magnetic field. This is a standard approach in superconducting electronics \cite{Likharev_1991,Tolpygo_2019,Mukhanov_2019}. A unique feature of our diodes is that they do not require an additional control line. This represents a significant advantage for complex and densely integrated circuits, alleviating the interconnect bottleneck problem.

A critical consideration for dense circuits is the degree to which the CL current is localized and the magnitude of cross-talk it may induce.  
A common strategy to reduce stray magnetic fields in superconducting circuits is the introduction of a ground plane, which confines the in-plane magnetic field to the narrow separation region between two films while cancelling it outside the structure. This approach is well suited for self-field diodes based on overlap-type JJs \cite{Krasnov_1997,Guarcello_2024}.
However, for planar JJs, a ground plane suppresses the out-of-plane magnetic field and thereby eliminates the first of the three self-field contributions discussed above, namely $L_g$. In contrast, the second contribution, $L_k$, remains unaffected, while the third contribution may even be enhanced.

The observed pronounced difference in the $I_c(H)$ patterns of D1 and D2 in Figs. \ref{fig:1}(e) and (f) clarifies the main contribution to $L_{sf}$. The $L_g$ part is determined by the geometries of notches, which are similar for D1 and D2. The $L_k$ part depends on the current density and scales as $L_k \propto W_x / W_z$. For D1, $W_x / W_z \simeq 1$, while for D2 it is $\simeq 10$. Therefore, $L_g\text{(D2)} \simeq L_g\text{(D1)}$, while $L_k\text{(D2)} \simeq 10 L_k\text{(D1)}$. Thus, the observed large difference between D1 and D2 indicates the dominance of the kinetic contribution to $L_{sf}$. Consequently, if required, a ground plane can be implemented in such diodes with minimal impact on their operation. Furthermore, since $L_k$ is not associated with a real magnetic field, it is not expected to generate significant cross-talk, even in the absence of a ground plane.

The most striking evidence for the unrestricted tunability of our ScDs is provided by the observation of a strong over-tilting of the $I_c(H)$ patterns (Fig.~3(d)), exceeding by a wide margin the self-field limit for conventional JJs \cite{Vasenko_1981}. As shown in Figs.~3(e) and (f), the overlap of the central $I_c(H)$ lobe with subsequent secondary lobes leads to reentrant superconductivity with increasing current. This was surprising since the possibility of over-tilting has not been predicted before \cite{Vasenko_1981,Monaco_2012,Guarcello_2024}.

The observed reentrance is distinct from reports of reentrant unconventional superconductivity induced by variations of some external parameter \cite{Zdravkov_2010,Knebl_2019}. 
In our case, it originates from the coexistence of three critical currents $I_{c1,2,3}$ under identical conditions, Fig.~3(f).

Multivalued critical currents have been reported before in phase-shifted long JJs \cite{Goldobin_2013,Hovhannisyan_2024} and are commonly observed in JJ arrays \cite{Modes_2000,Grebenchuk_2022}. However, such metastability is typically accompanied by hysteretic behavior. In contrast, the $I$--$V$s of our diodes are fully reversible.

The nonhysteretic reentrant $I$-$V$ enables new functionality. For example, it allows for a simple realization of a Gauss neuron for neuromorphic computing \cite{Klenov_2016,Beck_2020,Bolginov_2025}. The red line in Fig.~3(f) represents a Gaussian fit, 
indicating that this JJ is a compact, ready-to-use Gauss neuron.

Finally, we discuss the significance of the pursuit of higher nonreciprocity: does it need to be infinite? The answer depends on the specific application. For the construction of logic gates and related digital components, a modest $A \sim 2$--$4$ may be sufficient.

Infinite nonreciprocity is required in two cases. First, for the processing of small high-frequency signals, since $A = \infty$ eliminates the threshold for signal detection, as demonstrated in Fig.~2(c).

Second, a perfect ScD would solve a major problem with signal isolation in complex superconducting circuits. In semiconducting electronics, a single FET on a line can open/close it for signal transmission, owing to the large on/off ratio.  
Superconducting electronics lacks an equivalent switching element, leading to current leakage into multiple interconnections.
Experience from semiconducting electronics shows that on/off ratios $>10^4$ are required to suppress current leakage to acceptable levels in VLSI and ULSI circuits. Accordingly, reconfigurable and/or tunable ScDs with $A > 10^4$ would be indispensable for signal routing and mitigation of current leakage.  
The level of current isolation achieved here, $A > 10^3$, represents a crucial step towards this goal.

To conclude, we have demonstrated prototypes of tunable diodes based on a conventional niobium superconductor. We show that the multiterminal structure of such diodes enables both reconfigurability of the diode polarity and essentially unrestricted {\em in-situ} tunability, which together may enable transistor-like operation—the key missing element in superconducting electronics.
The split-current modes introduced here enable operation without an external magnetic field and provide additional functionality. These compact, technologically scalable, and robust ``geometrical'' diodes, suitable for practical applications, may enable the simple realization of superconducting logic and AI components, as well as signal routing and effective current isolation, thereby opening new avenues in digital and neuromorphic computation.

\section*{Methods}

{ \bf Device fabrication}. The diodes were fabricated from a single thin (70 nm) niobium film, deposited by DC magnetron sputtering. 
The electrodes were patterned by photolithography and reactive ion etching. The variable thickness bridge-type planar JJs were made by Ga$^+$ focused ion beam (FIB) etching (a single line cut). 
The junction linear current density depends on cut depth.  
For D1 and D2 devices nominal depths were 100 and 120 nm, respectively, 
leading to significantly smaller $J_c$ for D2. 

At each side of the JJ, narrow notches separating JJ electrodes were made. The self-field is generated in the notches and the narrow width $\sim 100~$nm of the notches increases the geometric self-field inductance. The quantitative estimation of geometric and kinetic contributions to $L_{sf}$ is provided in the Supplementary.  

To determine the optimization strategy, several batches were made, each containing several diodes with different junction parameters and electrode geometries. In total more than ten diodes were analyzed in this study (see the Supplementary). Some diodes were post-processed by FIB to modify electrode geometry. Such modification is seen as two dark rectangles (FIB-etched areas) in the SEM image of D2, Fig. 1(c).      

{\bf Experimental methods}. 
The measurements were performed in a closed-cycle cryostat. The magnetic field perpendicular to the film was supplied by a superconducting solenoid. 
All $I$-$V$s presented in this manuscript are non-hysteretic.
High-precision lock-in measurements, Fig. \ref{fig:2} (c), were performed 
using a 30 s averaging time.

{\bf Numerical modeling}.The numerical simulations of the $I_c(H)$ characteristics shown in Figs.~1(d) 2(a,e,f) and 3(b-d) were performed within the framework of the one-dimensional sine-Gordon model:

\begin{equation}
    \frac{\partial^2 \varphi}{\partial \bar{x}^2}
    - \frac{\partial^2 \varphi}{\partial \bar{t}^2}
    - \alpha \frac{\partial \varphi}{\partial \bar{t}}
    = \sin \varphi,
\end{equation}
where 
$\alpha$ is the quasiparticle damping parameter. 
The spatial coordinate $\bar{x}$ is normalized by 
$\lambda_J$, and time $\bar{t}$ by the inverse plasma frequency $\omega_p^{-1}$. Calculations were performed in the overdamped case, $\alpha = 1$.

The self-field back-action is introduced via the boundary conditions:

\begin{equation}
    \frac{\partial \varphi}{\partial x}\bigg|_{x=0, W_x} =
    \frac{2\pi}{W_x H_0}
    \left( H_e + H_{sf} + H_{\text{CL}} \right),
\end{equation}
where $H_e$ is the externally applied magnetic field. The self-field $H_{sf}$ arises from the asymmetric spatial distribution of the bias current $I_b^*(x)$. The corresponding effective self-flux can be estimated as

\begin{equation}
    \Phi_{sf} = L_{sf} I_b^* \simeq \frac{\lambda_J}{W_x} \frac{\Phi_0}{H_0} H_{sf},
\end{equation}
where the factor $\lambda_J/W_x$ reflects the exponential decay of flux at $x\sim \lambda_J$. The CL contribution is given by
$H_{\text{CL}} = {H_{sf}}/{2}\cdot{I_{\text{CL}}}/{I_b^*}$,
where the factor of 2 reflects that the CL current flows along only one edge of the junction. 

Because of the asymmetric current injection, the self-field affects only one edge of the JJ. Consequently, full compensation requires an external field $H_e = -H_{sf}/2$, as $H_e$ contributes symmetrically to both boundaries.

To simulate the dynamic rectification, Figs.~\ref{fig:2}(e,f), a time-dependent ac-bias current was applied in the form:

\begin{equation}
    I_b^* = I_{ac} \sin(\omega t),
\end{equation}
with $\omega = 0.01\,\omega_p$ (the low-frequency regime).

Additional simulation details are provided in the Supplementary Information.

\subsection*{Acknowledgements}

\subsection*{Author Contributions}

R.A.H. performed measurements, analyzed data, carried out numerical simulations and participated in writing the manuscript. T.G. fabricated samples. A.L. contributed to transport measurements. V.M.K. conceived the project and wrote the manuscript with input from all authors.

\subsection*{Data availability}
Authors confirm that all relevant data are included in the paper and its supplementary information files. Additional data can be provided upon reasonable request.

\subsection*{Competing Interests} The authors declare no competing interests.

\end{document}